\documentclass[
 reprint,
preprintnumbers,
 amsmath,amssymb,
 aps]{revtex4-1}
\usepackage{cancel}
\usepackage{xcolor}

\usepackage[normalem]{ulem}
\usepackage[colorlinks,pdfusetitle]{hyperref}
\usepackage{multirow}
\usepackage{graphicx}
\usepackage{dcolumn}
\usepackage{bm}
\usepackage{tikz}

\begin{document}

\preprint{FERMILAB-PUB-24-0762-T}


\title{
Exploring the Interference between the\\ Atmospheric and Solar Neutrino Oscillation Sub-Amplitudes 
}

\author{Gabriela Barenboim}
\email{gabriela.barenboim@uv.es; \# orcid:  0000-0002-3249-7467}
\affiliation{Departament de F{\'i}sica Te{\'o}rica and IFIC, \href{https://ror.org/043nxc105}{Universitat de Val{\`e}ncia-CSIC}, E-46100, Burjassot, Spain}

 \author{Stephen J. Parke}
\email{parke@fnal.gov; \# orcid: 0000-0003-2028-6782}
\affiliation{Theoretical Physics Department, \href{https://ror.org/020hgte69}{Fermi National Accelerator Laboratory}, Batavia, IL 60510, USA}

\date{\today}

\begin{abstract}
The interference between the atmospheric and solar neutrino oscillation sub-amplitudes is said to be responsible for CP violation (CPV) in neutrino appearance channels. More precisely,  CPV is generated by the interference between the parts of the neutrino oscillation amplitude which are CP even and CP odd: even or odd when the neutrino mixing matrix is replaced with its complex conjugate.
 This is the CPV interference term, as it gives a contribution to the oscillation probability, the square of the amplitude, which is opposite in sign for neutrinos and anti-neutrinos and is unique. For this interference to be non-zero, at least two sub-amplitudes are required. There are, however, other interference terms, which are even under the above exchange, these are the CP conserving (CPC) interference terms. In this paper, we explore in detail these CPC  interference terms and show that they cannot be uniquely defined, as one can move pieces of the amplitude from the atmospheric sub-amplitude to the solar sub-amplitude  and vice versa.  This freedom allows one to move the CPC interference terms around, but does not let you eliminate them completely.  We also show that there is a reasonable definition of the atmospheric and solar sub-amplitudes for the appearance channels such that in neutrino disappearance probability there is no atmospheric-solar CPC interference term. However, with this choice, there is a CPC interference term within the atmospheric sector.  
\end{abstract}

\maketitle

\section{Introduction}
\label{sec:intro}

Neutrino oscillation physics is entering the precision era especially with JUNO~\cite{JUNO:2015zny}, Hyper-Kamiokande~\cite{Hyper-Kamiokande:2018ofw} and DUNE~\cite{DUNE:2020ypp} coming on line within the next decade.  To achieve high precision, high statistics experiments with small systematic uncertainties are required. The above experiments will measure the parameters of the neutrino masses and mixings with unprecedented precision and  can also be used to explore new physics scenarios~\cite{Arguelles:2022tki} and to stress test the three neutrino mixing scenario in new ways, see for example~\cite{SajjadAthar:2021prg}.\\

One of the proposals in the literature ~\cite{Huber:2019frh} is to use the JUNO experiment to explore the interference between the atmospheric and solar sub-amplitudes to stress test the three neutrino paradigm. After defining the atmospheric sub-amplitude in a specific way, they show that in $\bar{\nu}_e$ disappearance, as will be explored by JUNO, there is a CP conserving (CPC) interference term between the atmospheric and solar sub-amplitudes
that can be measured or constrained at the 4 $\sigma$ level by this experiment.\\

In this paper we explore, for neutrino appearance, whether the full amplitude can be split up into an atmospheric and solar sub-amplitudes that makes unique physics sense or is there some ambiguity in how one makes this separation.  We find that some separations between the atmospheric and solar sub-amplitudes make more physics sense than others, and there is at least a one parameter set of such separation that is  reasonable. This set includes the most common ones and includes the choice that was made in~\cite{Huber:2019frh}.   As we will show, as one varies this parameter, one moves part of the atmospheric sub-amplitude to the solar sub-amplitude and vice versa.   This affects the CPC interference term between the atmospheric-solar sub-amplitudes. For a special choice of this parameter, we show that there is no atmospheric-solar CPC interference term in neutrino disappearance but there is a CPC interference term within the atmospheric sector. We also argue that one cannot eliminate the CPC interference terms completely and that these interference terms are of particular significance if the unitarity triangle for the channel under discussion is an obtuse triangle, i.e. has one angle $> \pi/2$. \\

The outline of this paper is as follows: in the next section \ref{sec:nuapp}, we write the neutrino appearance, oscillation amplitude in a new way that demonstrates an intrinsic freedom in defining the atmospheric and solar oscillation sub-amplitudes and connects this to the standard neutrino appearance probability. In section \ref{sec:nupmns}, we use the standard form of the neutrino mixing matrix to demonstrate this freedom and show how the
appearance oscillation amplitudes can be written in an informative way.  In section \ref{sec:nudis}, we show how in the disappearance probabilities one can use this freedom to trade an atmospheric-solar CPC interference term for an CPC interference solely within the atmospheric sector.  In section \ref{sec:TV} we consider separating the amplitude into time reversal even and odd parts.  We close with a discussion and summary section \ref{sec:end}. In the appendices  \ref{appx:Amps}  \&  \ref{appx:ids} , we give some other alternative ways to write the neutrino oscillation amplitudeand some useful identities.

\section{Neutrino Appearance }
\label{sec:nuapp}

\subsection{The Amplitude}
\label{sec:nuamp}

For neutrino appearance, the oscillation amplitude is given by~\cite{Bilenky:1987ty} 
\begin{align}
{\cal A} &= W_1 e^{-im^2_1 L/2E} +W_2 e^{-im^2_2  L/2E} +W_3 e^{-im^2_3  L/2E}
\label{eq:amp1} \\[2mm]
&\text{where} \quad W_i \equiv U^*_{\alpha i}  U_{\beta i} \quad \text{with} \quad \alpha \neq \beta \, .
\notag
\end{align}
We suppress the $\alpha \beta$
channel designation, unless required for clarity.
U is the the Pontecorvo-Maki-Nakagawa-Sakata (PMNS) mixing matrix and $m^2_i$ are the neutrino masses squared. $L$ is the baseline and $E$ is the neutrino energy.\\

One could use the unitarity condition, 
\begin{align}
   W_1+W_2+W_3=0 \,, \label{eq:unitarity} 
\end{align} 
to eliminate any one of the three W's, however  we chose to rewrite this amplitude by repeatedly using the sum and difference of terms, with $\Delta_{ij} \equiv \Delta m^2_{ij}L/4E$, as
\begin{align}
{\cal A} &= W_3( \sin \Delta_{31} e^{-i\Delta_{32}} + \sin \Delta_{32} e^{-i\Delta_{31}}) \notag \\
&  \quad +(W_2- W_1)\sin \Delta_{21}  \,.
\label{eq:amp1+}
\end{align}
We have removed an over all phase, and without loss of generality, we assume $W_3$ is real. This maintains the requirement that all kinematic factors which involve $m^2_3$ are proportional to $W_3$, as we are interested in separating this full amplitude into an atmospheric, (31+32),  and a solar (21) sub-amplitude.\\

 However, since
\begin{align}
\sin \Delta_{31} e^{-i\Delta_{32}} -  \sin \Delta_{32} e^{-i\Delta_{31}} - \sin \Delta_{21}=0,
\label{eq:Kin_ID}
\end{align}
one can add $(\xi \, W_3)$ times this kinematic factor to the amplitude, where $\xi$ is an arbitrary complex variable. Then the oscillation amplitude is given by
\begin{align}
{\cal A}&=  W_3\, (\,(1+\xi)\sin \Delta_{31} e^{-i\Delta_{32}}+ (1-\xi)\sin \Delta_{32} e^{-i\Delta_{31}} \,)\notag \\
& \quad +(\, W_2-W_1 -\xi W_3\,) \sin \Delta_{21} \,.
\label{eq:amp2}
\end{align}
This amplitude satisfies 
$1 \leftrightarrow 2$  with $\xi \rightarrow -\xi$ symmetry for any value of $\xi$, and since $\xi$ is arbitrary, no physical observable can depend on this variable. However, it maybe that a particular choices for variable $\xi$ can help understand the physics better.\\

If $\xi=-1$ (+1), this is equivalent to eliminating the $W_1$ $(W_2)$ term using the unitarity condition in eq.~\ref{eq:amp1}. Whereas setting $\xi = (W_2-W_1)/W_3$ is equivalent to eliminating the $W_3$ term.
So this way of writing the amplitude, eq.~\ref{eq:amp2}, is more general than other ways but includes the elimination of one of the mass eigenstates with appropriate choice of  variable $\xi$. In appendix \ref{appx:Amps}, we have given some other, more explicit examples.  \\

To impose the constraint such that terms with $m^2_3$ be explicitly proportional to $W_3$, i.e. they vanish if $W_3$ is set to zero, we require that
\begin{align}
 \xi W_3 \rightarrow 0 \quad \text{ when}  \quad W_3 \rightarrow 0 \,.
 \label{eq:constr}
 \end{align}
  This eliminates the possibility of choosing $\xi$ such that
 \begin{align}
  \xi &= (W_2-W_1)/W_3 \,, \notag  \\ \text{or} \quad  \xi &= \Re[(W_2-W_1)/W_3] 
   \notag  \,, \\ \text{or} \quad
   \xi &= i\Im[(W_2-W_1)/W_3] \,,
 \end{align}
 i.e. all or  part of the (21) sub-amplitude is moved into (31+32) sub-amplitudes.
 We impose this constraint because we are interested in the interference between the atmospheric and solar sub-amplitudes. Note that   $W_3$ and $(W_2-W_1)$ are independent variables. 
  The variable $\xi$ is arbitrary at this point, up to the constraint imposed in eq.~\ref{eq:constr} and for the rest of the paper we will assume $\xi $ to be real. Even though, we are primarily interested in the interval  $[-1,+1]$ for $\xi$, in general, we have a real one-parameter family of oscillation amplitudes defined by eq.~ \ref{eq:amp2}.  \\

  Kinematically it is natural to identify the 31, 32 and 21 sub-amplitudes as
  \begin{align}
{\cal A}_{31}  &\equiv W_3  (1+\xi)\sin \Delta_{31} e^{-i\Delta_{32}} \,, \notag \\
{\cal A}_{32}  &\equiv W_3  (1-\xi)\sin \Delta_{32} e^{-i\Delta_{31}} \,,
\label{eq:31+32+21} \\
{\cal A}_{21}  &\equiv  ( W_2-W_1-\xi W_3 )\sin \Delta_{21} \,. \notag
\end{align}
In ${\cal A}_{31}$ and  ${\cal A}_{31}$, the complex parts of the sub-amplitudes are in the kinematic factors whereas for ${\cal A}_{21}$, the complex parts are in the coefficient in front of the kinematic factor.  This is an arbitrary but convenient choice, and any physics conclusion is independent of this choice.\\

Then, the atmospheric, ${\cal A}_{\oplus}$, and solar, ${\cal A}_{\odot}$, sub-amplitudes are naturally given by
 \begin{align}
{\cal A}_{\oplus}  &\equiv {\cal A}_{31}  + {\cal A}_{32}  \,, \notag  \\
{\cal A}_{\odot}  &\equiv  {\cal A}_{21} \,.
\label{eq:atmos+solar}
\end{align}
In the limit $m^2_3 \rightarrow m^2_2$, ${\cal A}_{\oplus}  \sim \sin \Delta_{21}$ and similarly for  $m^2_3 \rightarrow m^2_1$, so that $\nu_1$ and $\nu_2$ are treated symmetrically.  
When $m^2_2 \rightarrow m^2_1$, ${\cal A}_{\oplus}  \rightarrow 2W_3 \sin \Delta_{31}$ and ${\cal A}_{\odot} \rightarrow 0 $ which is natural in this limit. When $W_3 \rightarrow 0$, then ${\cal A}_{\oplus} \rightarrow 0$ and ${\cal A}_{\odot} \rightarrow 2W_2 \sin \Delta_{21}$ as expected (in this limit $W_1=-W_2$). Therefore, we consider the definitions given in eq.~\ref{eq:atmos+solar} to be very natural definitions for the atmospheric and solar sub-amplitudes. However, this identification only makes physics sense if one can find a $\xi$ such that the  sub-amplitudes  have an appropriate dependence on the mixing parameters or elements of the PMNS mixing matrix. It is not obvious that this can be done for all appearance channels, especially for the 21 sub-amplitude  ${\cal A}_{21} $, as it is given as the difference of two terms.\\

The square of the solar and atmospheric sub-amplitudes  are given by
\begin{align}
P_{\odot} \equiv |{\cal A}_{\odot}|^2&=  | W_2-  W_1-\xi W_3 |^2 \sin^2 \Delta_{21} \notag \\
P_{\oplus} \equiv |{\cal A}_{\oplus}|^2&=  |W_3|^2[(1+\xi)^2 \sin^2 \Delta_{31} + (1-\xi)^2 \sin^2 \Delta_{32} ]   \notag   \\&+ Q^{CPC}_{31/32} \,, \notag 
 \end{align}
 where the square of the atmospheric sub-amplitude contains a CPC interference term between the 31 and 32 sub-amplitudes. This interference term 
 is given by
 \begin{align}
Q^{CPC}_{31/32}&  \equiv (1-\xi^2)|W_3|^2[\sin^2 \Delta_{31}+\sin^2 \Delta_{32}-\sin^2 \Delta_{21}] \notag \,.
 \end{align}
 Including this interference, the square of the atmospheric sub-amplitude is  given by
 \begin{align}
P_{\oplus} \equiv |{\cal A}_{\oplus}|^2&=  |W_3|^2[2(1+\xi) \sin^2 \Delta_{31} + 2(1-\xi) \sin^2 \Delta_{32}
\notag \\
& -(1-\xi^2) \sin^2 \Delta_{21} ]   \,.
 \end{align}

 The CPC interference term between atmospheric and solar sub-amplitudes is given by
  \begin{align}
Q^{CPC}_{31+32/21}&  \equiv 2 \Re[W_3((1+\xi) W^*_2- (1-\xi) W^*_1)]
\notag \\
&\times [(1+\xi)   \sin \Delta_{31} \cos \Delta_{32}\sin \Delta_{21} \notag \\ 
&~~+(1-\xi)   \cos \Delta_{31} \sin\Delta_{32}\sin \Delta_{21}] \,, \notag \\[2mm]
& =2 \Re[W_3((1+\xi) W^*_2- (1-\xi) W^*_1)] \notag \\
& \times  ( \sin^2 \Delta_{31}-\sin^2 \Delta_{32} +\xi \sin^2 \Delta_{21}) \,.
\label{eq:Int_CPC}
\end{align}

 Whereas the CP violating (CPV) interference term ($Q^{CPV}_{ij/kl}$) is
 \begin{align}
 Q^{CPV}_{31+32/21}  &=8 J
\sin \Delta_{31}\sin \Delta_{32}\sin \Delta_{21} \,,
 \end{align}
where $J \equiv \Im[W_3W^*_2] =-\Im[W_3W^*_1] $ which is the Jarlskog invariant~\cite{Jarlskog:1985ht}. Only the CPV interference term is independent of the arbitrary variable $\xi$ and is therefore truly unique. This term is the interference between the odd and even parts of the amplitude when $U$  is replaced by $U^*$, as can be seen by the following:
\begin{align}
\text{If one writes} \quad {\cal A}(U) &={\cal A}_+(U) + {\cal A}_-(U) \notag \\
\text{with} \quad {\cal A}_\pm(U) &\equiv \frac1{2}({\cal A}(U) \pm {\cal A}(U^*) ) \notag \\
\text{then} \quad  |{\cal A}(U)|^2 &= |{\cal A}_+(U)|^2+ |{\cal A}_-(U)|^2+Q^{CPV}_{31+32/21} \notag \\ 
\text{where} \quad Q^{CPV}_{31+32/21}  &\equiv 2 \Re[{\cal A}^*_+ (U) {\cal A}_-(U)] \,.
\label{eq:CP}
\end{align}
This $Q^{CPV}_{31+32/21} $ is the only term in the square of the amplitude that is odd under $U \leftrightarrow U^*$ as  
${\cal A}_-$ changes sign whereas ${\cal A}_+$, $|{\cal A}_+|^2$ and $|{\cal A}_-|^2$ are all even. In vacuum, the amplitude ${\cal A}(U^*)$ can be considered to be either the anti-neutrino propagation amplitude or the neutrino amplitude with opposite sign of the Jarlskog invariant.  This paragraph is all that we will say about the CPV interference terms as this is all well known in the literature, see for example~\cite{Dick:1999ed}. In section \ref{sec:TV}, we relate this to T odd and T even parts of the probability. \\

  What about the
the CPC interference term, eq.~\ref{eq:Int_CPC} ? Is there a value of $\xi$ that makes more physics sense than other values ?
Clearly there are simple values, $\xi=0,\pm1$, but what about other possibilities ? We will explore these possibilities in the next sections.

\subsection{Via the Probability}
\label{sec:nuprob}

The kinematic form of the sub-amplitudes is given by
\begin{align}
\sin \Delta_{31} e^{-i\Delta_{32}}, \quad \sin \Delta_{32} e^{-i\Delta_{31}}, \quad \text{and} \quad \sin \Delta_{21}
\end{align}
Thus the CPC interference terms between these sub-amplitudes have the following kinematic form: \begin{align}
32/21: \quad & 2\cos \Delta_{31} \sin \Delta_{32} \sin \Delta_{21} \notag \\
&=\sin^2 \Delta_{31}- \sin^2 \Delta_{32}- \sin^2 \Delta_{21} \notag  \\
31/21: \quad & 2\sin \Delta_{31} \cos \Delta_{32} \sin \Delta_{21} \\
& =\sin^2 \Delta_{31}- \sin^2 \Delta_{32}+ \sin^2 \Delta_{21}  \notag  \\
31/32: \quad & 2\sin \Delta_{31} \sin \Delta_{32} \cos \Delta_{21} \notag \\
& =\sin^2 \Delta_{31}+ \sin^2 \Delta_{32}- \sin^2 \Delta_{21}  \notag 
\end{align}
Note that the kinematic signature of the interference terms for both 31/21 and 32/21 is $\sin^2 \Delta_{31}- \sin^2 \Delta_{32}$. This suggests that in the general form of the oscillation probabilities, we add and subtract the $ \sin^2 \Delta_{3i}$ pieces, as follows,
\begin{align}
P\equiv |{\cal A}|^2 &=- 4\sum_{i>j}   R_{ij}\, \sin^2 \Delta_{ij} +P_{CPV} \label{eq:prob}  \\
 =& -(R_{31}+R_{32})(2 \sin^2 \Delta_{31}+ 2 \sin^2 \Delta_{32}) \notag \\
 & - 2(R_{31}-R_{32})(\sin^2 \Delta_{31}- \sin^2 \Delta_{32}) \notag \\
 & -4R_{21} \sin^2\Delta_{21} +P_{CPV}  \,, \notag \\[2mm]
 =&~~ |W_3|^2 (2 \sin^2 \Delta_{31}+ 2 \sin^2 \Delta_{32}) \notag\\
 &+ 2\Re[W_3(W^*_2-W^*_1)] \sin ( \Delta_{31} + \Delta_{32} )\sin\Delta_{21} \notag \\
 & +(|W_2-W_1|^2 -|W_3|^2)\sin^2 \Delta_{21} +P_{CPV}\,, \notag
 \end{align}
 where $P_{CPV} \equiv Q^{CPV}_{31+32/21}$.  See appendix~\ref{appx:ids} for some useful identities.
  After combining the terms proportional to $|W_3|^2$, it is clear that the amplitude can be written as before
 \begin{align}
 {\cal A}=& W_3(\sin \Delta_{31} e^{-i\Delta_{32}}+\sin \Delta_{32} e^{-i\Delta_{31}})  \notag \\ & \quad +(W_2-W_1)\sin \Delta_{21}   \notag \\
 \text{or}\quad \quad  & \notag\\
 =&  W_3(\sin( \Delta_{31}+ \Delta_{32} )-2i \sin \Delta_{31} \sin \Delta_{32}  )  \notag \\ & \quad +(W_2-W_1)\sin \Delta_{21}  \,.
 \label{eq:amp_ap}
\end{align}
Although this is aesthetically pleasing, because of the simple $1 \leftrightarrow 2 $ interchange symmetry, as we have seen in the previous section, we can add a term proportional to the identity given in eq.~\ref{eq:Kin_ID} to the amplitude. If this term is added, as was done for eq.~\ref{eq:amp2}, then the coefficients of the $\sin^2 \Delta_{ij}$ in eq.~\ref{eq:prob} are given as
 \begin{align}
  -4R_{31} &=(1+\xi)^2 |W_3|^2 +  (1-\xi^2)  |W_3|^2  \notag \\ &\hspace*{1.5cm}
  +2 \Re[W_3( W^*_2-W^*_1-\xi W^*_3)] \label{eq:R31} \\
   & \hspace*{-0.cm} =2(1+\xi) |W_3|^2  
 +2 \Re[W_3( W^*_2-W^*_1-\xi W^*_3)]  \notag  \\
  & = 2|W_3|^2+2 \Re[W_3( W^*_2-W^*_1)]  \notag \,, \\[3mm]
  %
    -4R_{32}&=(1-\xi)^2 |W_3|^2 + (1-\xi^2)  |W_3|^2 \notag 
\\ &\hspace*{1.5cm}
   -2\Re[W_3( W^*_2-W^*_1-\xi W^*_3)]  \label{eq:R32}  \\
  &=2(1-\xi) |W_3|^2   
 -2 \Re[W_3( W^*_2-W^*_1-\xi W^*_3)] \notag  \\
  & = 2|W_3|^2-2 \Re[W_3( W^*_2-W^*_1)]  \notag  \,, \\[3mm]
  %
  -4R_{21}&= | W_2-W_1-\xi W_3 |^2 
 - (1-\xi^2) |W_3|^2 \notag  \\ &
\hspace*{1.5cm} + 2~ \Re[~\xi W_3( W^*_2-W^*_1-\xi W^*_3)~] \label{eq:R21}   \\
 &=  | W_2-W_1 |^2 - |W_3 |^2  \notag  \,. 
 \end{align}
Here, for each $R_{ij}$, the first term is $|{\cal A}_{ij}|^2$, the second is the CPC interference between 31/32, and the third is CPC interference between (31+32)/21. Only the first term is guaranteed to be positive and the sum must be independent of the variable $\xi$, as is shown on the final line of each $R_{ij}$. \\
 
\begin{center}
        \begin{figure}[!t]
           \includegraphics[width=0.5\textwidth]{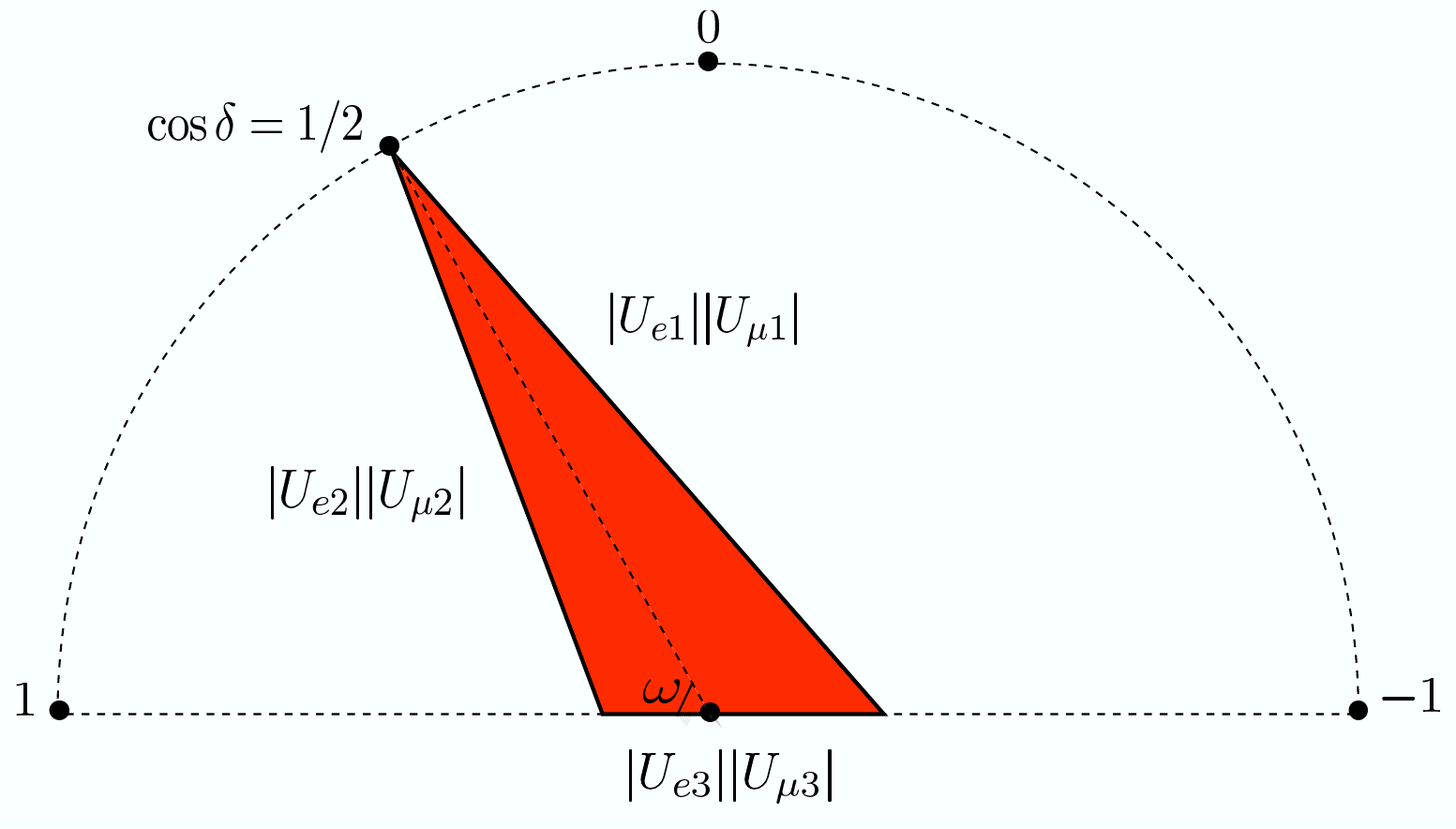}
        \caption{The $e\mu$ unitarity triangle, where $|W_i| \equiv |U_{e i}||U_{\mu i}| $ As one varies the cosine of the CP violating phase, $\cos \delta$, from 1 to -1, this triangle goes from obtuse, to acute near $\cos \delta = 0$ or maximal CP violation, back to obtuse.  All the $R_{ij}$'s are negative only when the triangle is acute which occurs near maximal CP violation or $\cos \delta$ near zero. When this triangle is  a obtuse (right) triangle one of the $R_{ij}$ is positive (zero). This figure uses the PMNS matrix from eq.~\ref{eq:PMNS} and is adapted from~\cite{Nunokawa:2007qh}.          }      
            \label{fig:Triplot}
        \end{figure}
    \end{center}

 As $W_1, W_2, W_3$ are the sides of a unitarity triangle and
 \begin{align}
 2R_{ij} &= |W_k|^2-|W_j|^2 -|W_i|^2 \,,
 \end{align}
with (i,j,k) all different, there are three possibilities:
 \begin{enumerate}
 \item if the unitarity triangle is acute (all angles $< \pi/2$) then all $R_{ij}$'s are negative,
 \item if the angle between $W_i$ and $W_j$ is a right angle then $R_{ij}=0$ with the other two negative, 
 \item  if the angle between $W_i$ and $W_j$ is $ >\pi/2$, an obtuse triangle, then $R_{ij}>0$ with the others negative. This $R_{ij}>0$ is dominated by the CPC interference terms.
 \end{enumerate}
 See Fig.~\ref{fig:Triplot}.
 If an $R_{ij}$ is positive, then the dominant term that contributes to this must be a CPC interference term.
 As we will see in the next section, typically one of the  $R_{ij}>0$ for all appearance channels except when CPV is close to maximal, i.e. $|J|$ is close to its maximum value given the known other mixing parameters. 
 
 Therefore, from the length of the side of the channel's unitarity triangle one can calculate all of the $R_{ij}$'s for that channel, as well as the magnitude of the  Jarlskog invariant as the Jarlskog invariant is equal to half the area of the triangle. The area of a triangle can be calculated from the length of the sides, see~\cite{Nunokawa:2007qh} for more detail. The sign of $J$, cannot be calculated, as the lengths of the sides are invariant under the CP interchange. \\

\section{The Neutrino Appearance Amplitudes: Specifics}
\label{sec:nupmns}

\begin{center}
        \begin{figure}[!t]
           \includegraphics[width=.41\textwidth]{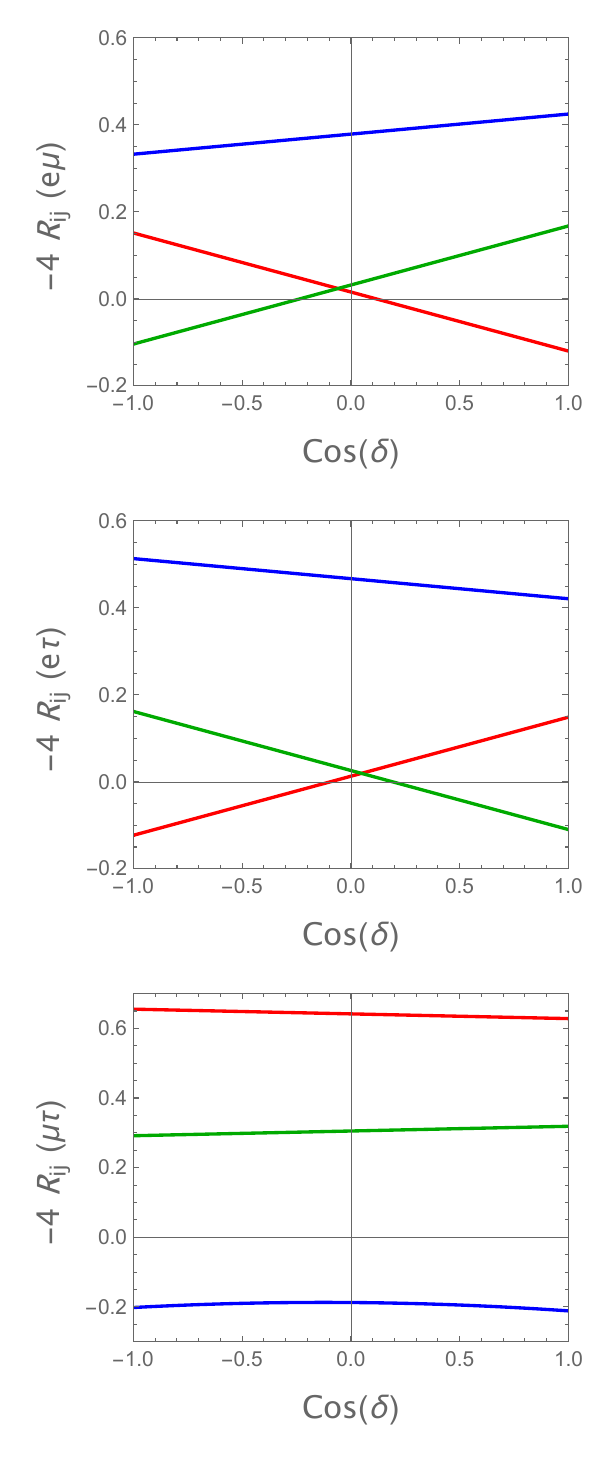} \\          
          \includegraphics[width=.35\textwidth]{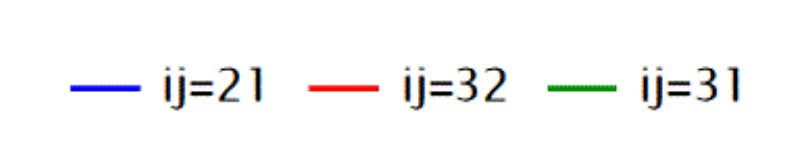}
        \caption{The minus four R's, $(-4R_{ij})$, for the channels $e\mu$, $e\tau$ and $\mu \tau$ as a function of $\cos \delta$. For the lines that are negative, the dominant contribution is from a CPC interference term. For a fixed set of parameters, the following must be satisfied:  in any one channel, at most one $-4R_{ij}$ can be negative and any $-4R_{ij}$ can be negative in at most one channel, see comment in appendix~\ref{appx:ids}. These constrains are satisfied in the above figures. For this figure we used $(s^2_{13}, s^2_{12}, s^2_{23} )=(0.022, 0.33,0.55)$.}      
            \label{fig:Rplot}
        \end{figure}
    \end{center}

Here we will use a slight modification of the PMNS matrix used by the PDG~\cite{ParticleDataGroup:2024cfk}, given by
{\small
 \begin{align}
&U=  \label{eq:PMNS} 
 \\
 & \hspace{-0.1cm}\left( \begin{array}{ccc}
c_{13} c_{12} & c_{13}s_{12} & s_{13} \\
-s_{12} c_{23} e^{-i\delta} -s_{13} c_{12} s_{23} &c_{12} c_{23} e^{-i\delta} -s_{13} s_{12} s_{23} & c_{13} s_{23} \\
s_{12} s_{23} e^{-i\delta} -s_{13} c_{12} c_{23} &-c_{12} s_{23} e^{-i\delta} -s_{13} s_{12} c_{23} & c_{13} c_{23} \\
\end{array}
\right) .  \notag
\end{align}
}
This choice minimizes the number of complex elements in U. In fig.~\ref{fig:Rplot}, we show the various $R_{ij}$'s for the 3 appearance channels as functions of $\cos \delta$, the CP phase. Different choices of $\xi$ can illuminate the physics of these plots.
\\

\hspace{1cm}

For the $e\mu$ channel, calculation of the $W^{e\mu}_i$'s gives
 \begin{align}
\left( \begin{array}{c}W^{e\mu}_1 \\ W^{e\mu}_2 \\W^{e\mu}_3\\\end{array}\right)
=  \left( \begin{array}{rcl}
-c_{23} c_{13}  s_{12}c_{12}  e^{-i\delta} &-& s_{23} s_{13} c_{13}  \, (c^2_{12}) \\[2mm]
 ~c_{23} c_{13}  s_{12} c_{12}  e^{-i\delta} &-& s_{23} s_{13} c_{13}  \, (s^2_{12}) \\[2mm]
 &&s_{23} s_{13} c_{13}  \\
\end{array}
\right) , \label{eq:Wem}
\end{align}
where the combination $s_{23} c_{23}$ does not appear.
Now if we set $\theta_{23}=0$,  we have only a non-zero 21 sub-amplitude. Whereas if we set $\theta_{23}=\pi/2$, only the sub-amplitudes 31 and 32 are non-zero~\footnote{One could also phrase this in a PMNS parameterization independent way; if $U_{\mu 3}=0$ then we have purely 21 oscillations and if $U_{\tau 3}= 0$, we have no 21 oscillations, only 31 and 32 oscillations.}. In this limit, $R^{e\mu}_{21}>0$ comes purely from the CPC interference of the 31 and 32 sub-amplitudes. Clearly an informative choice for $\xi$ is
\begin{align}
\xi &= c^2_{12}-s^2_{12} \,,
\label{eq:xiemu}
\end{align}
then
\begin{align}
A^{e\mu}_{21} \equiv (W^{e\mu}_2 - W^{e\mu}_1- \xi W^{e\mu}_3) & = 2 c_{23} c_{13}  s_{12} c_{12}  e^{-i\delta} \,, \notag \\
A^{e\mu}_{31} \equiv (1+\xi) W^{e\mu}_3 &= 2 c^2_{12} \, s_{23} s_{13} c_{13}  \,,  \notag \\
A^{e\mu}_{32} \equiv (1-\xi)W^{e\mu}_3 &= 2 s^2_{12} \, s_{23} s_{13} c_{13} \,.
\end{align}
Therefore, the sub-amplitudes are given by~\cite{Parke:2018shx}
 \begin{align}
 {\cal A}^{e \mu}_{31} &=2s_{13}c_{13}s_{23} c^2_{12} \sin \Delta_{31} e^{-i \Delta_{32}} \,, \notag\\
 {\cal A}^{e \mu}_{32} &=2s_{13}c_{13}s_{23} s^2_{12} \sin \Delta_{32} e^{-i \Delta_{31}} \,, \label{eq:Ampem}\\
 {\cal A}^{e \mu}_{21} &= 2 s_{12} c_{12} c_{13}  c_{23} e^{-i \delta} \sin \Delta_{21} \,. \notag
 \end{align}
 The physical interpretation of this form of the $e \mu$  amplitude is that when $s_{23}=0$, the $e\mu$ channel is purely a 21 oscillation, and when  $s_{23}=1$ there are only 31 and 32 oscillations, in between, all oscillations are present. With this interpretation, the CPC interference between the solar and atmospheric oscillations is proportional to $s_{23} c_{23}$, and is given by
  \begin{align}
 Q^{CPC}_{31+32/21}(\nu_e \rightarrow \nu_\mu)&= 4 s_{23}c_{23} \, s_{13}c^2_{13} \, s_{12} c_{12} \cos \delta \notag  \\
 & \hspace{-3 cm} \times (\sin^2 \Delta_{31}-\sin^2 \Delta_{32}+(c^2_{12}-s^2_{12})\sin^2 \Delta_{21} ) \,.
 \label{eq:Qme}
 \end{align} 
 The coefficient of $\cos \delta$ here, $~4 s_{23}c_{23} \, s_{13}c^2_{13} \, s_{12} c_{12}$, ~gives the slope of the three lines for the $e \mu$ channel in fig.~\ref{fig:Rplot} when multiplied by $\pm1$ for $R_{31/32}$ and $(c^2_{12}-s^2_{12})$ for $R_{21}$. These are the last terms in eq.~\ref{eq:R31}-\ref{eq:R21}.  The offsets from passing through the origin are exactly given by 
 $2(1+\xi)|W_3|^2 = 4c^2_{12} s^2_{13} c^2_{13} s^2_{23} $ and $2(1-\xi)|W_3|^2 = 4s^2_{12} s^2_{13} c^2_{13} s^2_{23} $ for 31 and 32  respectively. For 21 the approximate offset is
 $ |W_2-W_1-\xi W_3|^2 =  4 c^2_{13} s^2_{12} c^2_{12} c^2_{23}$ with an additional term 
 $-(1-\xi^2) |W_3|^2 \sim {\cal O}(s^2_{13})$.

   In summary, for the $e \mu$ channel, using $\xi$ from eq.~\ref{eq:xiemu},  then $R_{31}$ and $R_{32}$ are dominated by the atmospheric-solar interference. However, their sum is given by $R_{31}+R_{32}=-|W_3|^2$, so both cannot be positive at the same value of $\cos \delta$.  For $R_{21}$ is dominated by the square of the 21 sub-amplitude for all  $\cos \delta$. \\

A similar argument for the $e\tau$ channel, using the same value of $\xi$, gives
\begin{align}
 {\cal A}^{e \tau}_{31} &=2s_{13}c_{13}c_{23} c^2_{12} \sin \Delta_{31} e^{-i \Delta_{32}} \,, \notag \\
 {\cal A}^{e \tau}_{32} &=2s_{13}c_{13}c_{23} s^2_{12} \sin \Delta_{32} e^{-i \Delta_{31}} \,, \label{eq:Ampet}\\
 {\cal A}^{e \tau}_{21} &= -2 s_{12} c_{12} c_{13}  s_{23} e^{-i \delta} \sin \Delta_{21} \,.  \notag
 \end{align}
Note that $e\tau$ amplitude is obtained from $e\mu$ amplitude by $s_{23} \rightarrow c_{23}$ plus $c_{23} \rightarrow -s_{23}$, as expected. The physical interpretation of this form of $e\tau$ amplitude is similar to $e\mu$ with the roles of $s_{23}$ and $c_{23}$ reversed. The CPC interference between the solar and atmospheric oscillations is again proportional to $s_{23} c_{23}$, and is equal and opposite in sign to the $e \mu$ channel:
 \begin{align}
 Q^{CPC}_{31+32/21}(\nu_e \rightarrow \nu_\tau)&= - 4 s_{23}c_{23} \, s_{13}c^2_{13} \, s_{12} c_{12} \cos \delta \notag  \\
 & \hspace{-3 cm} \times (\sin^2 \Delta_{31}-\sin^2 \Delta_{32}+(c^2_{12}-s^2_{12})\sin^2 \Delta_{21} ) \,.
  \label{eq:Qet}
 \end{align} 
 The implications for the slopes of the lines in fig.~\ref{fig:Rplot} for $e\tau$ channels is the same as that for $e\mu$ channels with opposite sign.
 As we will see in the next section, this equality but opposite sign has implications for the $\nu_e$ disappearance channel.
\\

\begin{widetext}
Now for the $\mu \tau$ channel, the W's are given by
\begin{align}
 \left( \begin{array}{c}W^{\mu\tau}_1 \\ W^{\mu\tau}_2 \\W^{\mu\tau}_3\\\end{array}\right)
 =& \left( \begin{array}{ccl}
 ~ \, \, \biggr(s_{13}s_{12} c_{12}(c^2_{23}e^{+i\delta}  -s^2_{23} e^{-i\delta}) +s^2_{13} s_{23}c_{23}(c^2_{12}-s^2_{12}) \biggr)  &- & s_{23}c_{23} c^2_{13}\, (s^2_{12})\\[2mm]
 -\biggr(s_{13}s_{12} c_{12}(c^2_{23}e^{+i\delta}  -s^2_{23} e^{-i\delta})
 +s^2_{13} s_{23}c_{23}(c^2_{12}-s^2_{12})\biggr) &- &s_{23}c_{23} c^2_{13}\, (c^2_{12}) \\[2mm]
0  &+&s_{23} c_{23} c^2_{13}
 \end{array}
\right) \,.  \label{eq:Wmt} 
\end{align}
\end{widetext}

For this channel, the ``sorting hat'' is not $\theta_{23}$ but $\theta_{13}$, as the combination $s_{13} c_{13}$ does not appear. Now if we set $c_{13}=0$, we have only a non-zero 21 sub-amplitude, whereas if we set $s_{13}=0$, we have only the sub-amplitudes 31 and 32 that are non-zero.
 Again,  in this limit $R^{\mu \tau}_{21}>0$ comes from the CPC interference of the 31 and 32 sub-amplitudes. Therefore, it is clear we should choose $\xi= s^2_{12} -c^2_{12}$ such that
\begin{align}
 W^{\mu \tau}_2 - W^{\mu \tau}_1 -\xi W^{\mu \tau}_3  &= \notag \\
 -2\biggr(s_{13}s_{12} c_{12}(c^2_{23}e^{+i\delta}  -s^2_{23} e^{-i\delta}) &+s^2_{13} s_{23}c_{23}(c^2_{12}-s^2_{12})\biggr) \,,  \notag \\
(1+\xi)W^{\mu \tau}_3 &=2 s_{23} c_{23} c^2_{13}  \, (s^2_{12}) \,,  \notag \\
(1-\xi)W^{\mu \tau}_3 &= 2 s_{23} c_{23} c^2_{13}  \, (c^2_{12})  \,.
\end{align}
For the $\mu \tau$ channel, $\xi$ has opposite sign to what was used for the $e\mu$ and $e\tau$ channels.

Therefore, the total amplitude for the $\mu \tau$ channel is 

\begin{align}
 {\cal A}^{\mu \tau}_{31} &=2s_{23}c_{23}c^2_{13} s^2_{12} \sin \Delta_{31} e^{-i \Delta_{32}} \,, \notag \\
 {\cal A}^{\mu \tau}_{32} &=2s_{23}c_{23}c^2_{13} c^2_{12} \sin \Delta_{32} e^{-i \Delta_{31}} \,, \label{eq:Ampmt}\\
 {\cal A}^{\mu \tau}_{21} &= -2\biggr(s_{13} s_{12} c_{12} (c^2_{23}e^{+i \delta}-s^2_{23}e^{-i \delta }) \notag \\& \hspace{2cm}  + s^2_{13} s_{23} c_{23}(c^2_{12}-s^2_{12}) \biggr) \sin \Delta_{21}  \notag \,.
 \end{align}
 
 The physical interpretation of this form of the $\mu \tau$  amplitude is that when $c_{13}=0$, the $\mu \tau$ channel is purely a 21 oscillation, and when  $s_{13}=0$, there are only 31 and 32 oscillations, in between, all oscillations are present. However, since ${\cal A}_{21}$ is proportional to $s_{13}$, the dominant contribution to $R_{21}$ for this channel comes from the interference between the 31 and 32 sub-amplitudes, as seen in Fig.~\ref{fig:Rplot}.   The three curves in Fig.~\ref{fig:Rplot} can be approximated by $2(1+\xi)|W_3|^2, 2(1-\xi)|W_3|^2$ and $ -(1-\xi^2)|W_3|^2$ with $\xi=s^2_{12}-c^2_{12}$, i.e. they are dominated by the square of the atmospheric sub-amplitude:
 
 \begin{align}
 -4R^{\mu \tau}_{31}& \approx 2(1+\xi)|W^{\mu \tau}_3|^2 = 4c^2_{12} \, c^2_{13} s^2_{23} c^2_{23} \,,
 \notag  \\
 -4R^{\mu \tau}_{32}& \approx 2(1-\xi)|W^{\mu \tau}_3|^2 = 4s^2_{12} \, c^2_{13} s^2_{23} c^2_{23} \,,
  \notag \\
 -4R^{\mu \tau}_{21}& \approx -(1-\xi^2)|W^{\mu \tau}_3|^2 = -4 s^2_{12} c^2_{12} \, c^2_{13} s^2_{23} c^2_{23} \,.
 \end{align}
 
 The atmospheric-solar CPC interference term is given by
\begin{align}
 &Q^{CPC}_{31+32/21}(\nu_\mu \rightarrow \nu_\tau) =
 \notag  \\
 &-4 \biggr[ ~ s_{23}c_{23} \, s_{13}c^2_{13} \, s_{12} c_{12} \cos \delta \cos 2\theta_{23}
  \,  \notag \\  & \hspace*{2cm}  
 +  s^2_{13} c^2_{13}  s^2_{23} c^2_{23} \cos 2\theta_{12} ~ \biggr] \notag  \\
 & \times (\sin^2 \Delta_{31}-\sin^2 \Delta_{32}-\cos 2\theta_{12} \sin^2 \Delta_{21} ) \,.
  \label{eq:Qmt}
 \end{align} 

The sub-amplitudes using the above choices for $\xi$ are summarized in Table~\ref{tab:ampsum}.

\begin{widetext}

\begin{center}

\begin{table}[!t]
\begin{tabular}{|c|c|c|c|}
\hline
&&&\\[-2mm]
& ${\cal A}_{31}=$ & ${\cal A}_{32}=$  & ${\cal A}_{21}=$\\[2mm]
\hline
&&&\\[-2mm]
& $ 2\sin \Delta_{31} e^{-i\Delta_{32}} \times$ &. $2\sin \Delta_{32} e^{-i\Delta_{31}} \times$  & $2\sin \Delta_{21} \times$\\[2mm]
 \hline
&&&\\[-3mm]
$\mu \rightarrow e$ & $s_{23} s_{13} c_{13} c^2_{12}$ & $s_{23} s_{13} c_{13} s^2_{12}$ & $c_{23} s_{12} c_{12} c_{13} e^{+i\delta}$ \\[2mm]
\hline
&&&\\[-3mm]
$\tau \rightarrow e$ & $c_{23} s_{13} c_{13} c^2_{12}$ & $c_{23} s_{13} c_{13} s^2_{12}$ & $-s_{23} s_{12} c_{12} c_{13} e^{+i\delta}$ \\[2mm]
\hline
&&&\\[-3mm]
$\mu \rightarrow \tau $ &  $s_{23} c_{23}  c^2_{13} s^2_{12}$ & $s_{23} c_{23}  c^2_{13} c^2_{12}$ 
& ~$-s_{13}s_{12}c_{12} (c^2_{23}e^{+i\delta}-s^2_{23}e^{-i\delta}) ~$  \\[2mm]
 &   & & $-s^2_{13}s_{23} c_{23}(c^2_{12}-s^2_{12})$  \\[2mm]
 \hline
\end{tabular}
\caption{Summary of the sub-amplitudes for the different channels. With this set of sub-amplitudes,  CPV comes from $\Im{ ({\cal A}_{31} +{\cal A}_{32}) } \times  \Im{({\cal A}_{21} )} \sim \pm J~\Pi_{i>j} \sin \Delta_{ij}$ 
.}
\label{tab:ampsum}
\end{table}
\end{center}

\end{widetext}

 \section{Neutrino Disappearance Channels}
 \label{sec:nudis}
 
 As we saw in the $e\mu$ and $e\tau$ channels, if we chose $\xi= c^2_{12}-s^2_{12}$ for both, then the CPC interference terms between the atmospheric and solar amplitudes are equal but opposite in sign 
 so that for $\nu_e$ disappearance these CPC interference terms cancel.  There remain CPC interference terms between the 31 and 32 sub-amplitudes.  This suggest writing the $\nu_e $ disappearance probability as 
 \begin{align}
 &1-P(\nu_e \rightarrow \nu_e) =  P(\nu_e \rightarrow \nu_\mu) +P(\nu_e \rightarrow \nu_\tau) \notag \\
 & =4 c^2_{13} s^2_{12} c^2_{12} \sin^2 \Delta_{21}\\
 & +4 s^2_{13} c^2_{13} (c^4_{12} \sin^2 \Delta_{31}+s^4_{12} \sin^2 \Delta_{32})  \notag \\
 & +4 s^2_{13} c^2_{13} s^2_{12} c^2_{12}(\sin^2 \Delta_{31}+\sin^2 \Delta_{32}-\sin^2 \Delta_{21})  \,,\notag 
 \end{align}
where the last line is the 31/32 CPC interference term. Combining terms with the same $\sin^2 \Delta_{ij}$ gives the standard result.
Including this  CPC interference term changes the $c^2_{13}$ to $c^4_{13}$ in the coefficient of the solar $\sin^2 \Delta_{21}$ term and replaces the $(c^4_{12},\,s^4_{12})$  with $(c^2_{12},\, s^2_{12})$ in the coefficients of the atmospheric  $(\sin^2 \Delta_{31}, \, \sin^2 \Delta_{32})$ terms. The effect on the solar term is about 2\% whereas the effect on the atmospheric terms is significantly larger, $>$20 \%. 
Therefore absence of this CPC interference term would have significant implications for all reactor disappearance experiments, especially JUNO.   \\

Can this cancelation of the atmospheric/solar CPC interference be extended to both $\nu_\mu$ and $\nu_\tau$ disappearance? The answer is yes, for  $\nu_\alpha$ disappearance, if one uses
 \begin{align}
 \xi_\alpha =   \frac{|U_{\alpha 1}|^2-|U_{\alpha 2}|^2 }{(1-|U_{\alpha 3}|^2) } \,,
 \label{eq:xi-alpha}
 \end{align}
 in both $\alpha \beta$ and $\alpha \gamma$ channels, $(\alpha, \beta, \gamma)$ all different.
 For $\nu_e$ disappearance, this is the same $\xi$ as used in the last section. For  $\nu_\mu$ disappearance, $\xi$ it is approximately the same as for $\mu \tau$ in the last section with corrections of ${\cal O}(s_{13})$, and similarly for $\nu_\tau$ disappearance.\\
 
 By explicit calculation one can show that for $\nu_\mu \rightarrow \nu_e$ and $\nu_\mu \rightarrow \nu_\tau$ that 
 \begin{align}
 &Q^{CPC}_{31+32/21}(\nu_\mu \rightarrow \nu_e)=-Q^{CPC}_{31+32/21}(\nu_\mu \rightarrow \nu_\tau)
 \notag  \\
 &=4 \biggr[ ~s_{23}c_{23} \, s_{13}c^2_{13} \, s_{12} c_{12} \cos \delta \, (c^2_{23}-s^2_{13}s^2_{23})  \notag \\
 & \quad \quad  +  s^2_{13} c^2_{13}  s^2_{23} c^2_{23} (c^2_{12}-s^2_{12})  ~ \biggr]  \biggr/
 (c^2_{23}+s^2_{13}s^2_{23})\notag  \\
 & \times (\sin^2 \Delta_{31}-\sin^2 \Delta_{32}+\xi_\mu \sin^2 \Delta_{21} )
 \end{align} 
which for the $\mu e$ channel differs by  ${\cal O}(s_{13})$ when compared to  eq.~\ref{eq:Qme}. For $\mu \tau$ channel the differences when compared to eq.~\ref{eq:Qmt} are more significant.
 
One can show the cancellation directly, without an explicit calculation, using eq.~\ref{eq:Int_CPC}. We have to add the two channels
 $\alpha \beta$ and $\alpha \gamma$, therefore we have
\begin{align}
& \Re[\,W^{\alpha \beta}_3 |U_{\alpha1}|^2 (W^{\alpha \beta}_2)^*\,] + \Re[\,W^{\alpha \gamma}_3 
|U_{\alpha 1}|^2 (W^{\alpha \gamma}_2)^*\,] \notag \\[2mm]
=& -|U_{\alpha 3}|^2 |U_{\alpha 1}|^2 |U_{\alpha 2}|^2  \,, \notag  \\
& \Re[\,W^{\alpha \beta}_3 |U_{\alpha 2}|^2 (W^{\alpha \beta}_1)^*\,] + \Re[\,W^{\alpha \gamma}_3 
|U_{\alpha 2}|^2 (W^{\alpha \gamma}_1)^*\,] \notag \\
=& -|U_{\alpha 3}|^2 |U_{\alpha 1}|^2 |U_{\alpha 2}|^2    \,,
\end{align}
giving zero when subtracted. Therefore, this cancellation can be achieved for all disappearance channels.
Note that, this proof is independent of the  form of the PMNS matrix used in the previous section.\\

Thus  the general disappearance probability can be written as follows:
\begin{align}
1-P(\nu_\alpha \rightarrow \nu_\alpha) &=P(\nu_\alpha \rightarrow \nu_\beta)+P(\nu_\alpha \rightarrow \nu_\gamma) \notag \\
& \hspace{-2cm}  =\frac{4 |U_{\alpha 1}|^2 |U_{\alpha 2}|^2 }{(1-|U_{\alpha 3}|^2) }\sin^2 \Delta_{21} \notag 
 \\[2mm] 
 &  \hspace{-2cm} +
 \frac{4 |U_{\alpha 3}|^2}{(1-|U_{\alpha 3}|^2) }  \biggr(|U_{\alpha 1}|^4 
\sin^2 \Delta_{31} +|U_{\alpha 2}|^4 \sin^2 \Delta_{32} \biggr) \notag \\[2mm]
&  \hspace{-2.5cm}  +
 \frac{4|U_{\alpha 1}|^2|U_{\alpha 2}|^2  |U_{\alpha 3}|^2 }{(1-|U_{\alpha 3}|^2) }   \biggr(\sin^2 \Delta_{31} +\sin^2 \Delta_{32} -\sin^2 \Delta_{21} \biggr) \,,
 \label{eq:disapp}
\end{align}
where the last line is the CPC interference between the 31/32 sub-amplitudes with no (31+32)/21 interference. Again, combining terms with the same $\sin^2 \Delta_{ij}$ gives the standard result.\\

If one choses $\xi=1$ ( or $\xi=-1$) there is no 32/21 ( 31/21 ) interference as the 32 (31) sub-amplitude is set to zero.  When comparing these two choices the size of the CPC interference between atmospheric and solar sub-amplitudes differ in size and also sign.   If one chooses $\xi$ as in eq.~\ref{eq:xi-alpha} then there is no (31+32)/21 interference but there is 31/32 interference. Clearly the CPC interference terms, are dependent on how one splits the amplitude into an atmospheric and solar sub-amplitudes. As we have demonstrated there is no unique way to make this split and different choices will give different results for the CPC interference terms.  

\section{Using Time Reversal}
\label{sec:TV}

In this section we use time reversal to show that the T odd or time reversal violating part of the oscillation probability is also unique. Starting with the amplitude as given in eq.~\ref{eq:amp2}, if we group terms according to whether they are even, ${\cal A}_E$, or odd, ${\cal A}_O$,  when $L \rightarrow -L$ we have
\begin{align}
 {\cal A}_E &\equiv -2i\, W_3 \sin \Delta_{31} \sin \Delta_{32} \notag \\
 {\cal A}_O & \equiv W_3( (1+\xi)  \sin\Delta_{31} \cos \Delta_{32} +(1-\xi)  \sin\Delta_{32} \cos \Delta_{31}) \notag  \\& +  (W_2-W_1-\xi W_3)\sin \Delta_{21} \,. \label{eq:Tamp} 
\end{align}
Note that the dependence on $\xi$ naturally drops out of the T even term and clearly there is a $1 \leftrightarrow 2$ plus $\xi \rightarrow -\xi$ symmetry. We have singled out $3$ so that all terms with $m^2_3$ have an explicit $W_3$ dependence as discussed earlier. By permuting $(1,2,3)$, one could have singled out $1$ or $2$, however $3$ is chosen because it is uniquely associated with the atmospheric oscillation. \\

The appearance probability is then given by
\begin{align}
    P & \equiv |{\cal A}_E|^2+|{\cal A}_O|^2 +2 \Re[{\cal A}_E {\cal A^*}_O] \,,
\end{align}
where only the last, interference term is T Violating and $\xi$ independent. However, trying to split the T conserving part of the oscillation probability into an atmospheric and solar amplitude squared and their interference is ambiguous as before with the variable $\xi$ providing a one parameter set of such divisions.\\

When discussing this issue for neutrinos propagating in matter, it maybe clearer for the reader to use the T even and odd parts, rather than CP even and odd parts because matter effects are different for neutrinos and anti-neutrinos propagating in matter, thereby inducing additional CP violation terms.

\section{Discussion and Conclusions}
\label{sec:end}

We have explored the CPC interference between the atmospheric and solar sub-amplitudes in detail for three flavor neutrino oscillations. For neutrino appearance amplitudes, we have shown that one can move parts of the amplitude from the atmospheric to the solar sub-amplitudes and vice versa, and there is no unique and therefore physical way to make this choice.  This implies that the CPC interference term between atmospheric and solar sub-amplitudes is also not unique, as one can move parts of this interference into a CPC interference within the atmospheric sector. In general, one cannot eliminate the CPC interference terms completely, but by choosing the sub-amplitudes in different ways one can move them around.  \\

This freedom has been  used  in eq.~\ref{eq:disapp} to force the CPC interference between the atmospheric and solar amplitudes in neutrino appearance for $\nu_\alpha \rightarrow \nu_\beta$ and  $\nu_\alpha \rightarrow \nu_\gamma$  ( with $\alpha, \beta, \gamma$ all different) to be equal in magnitude and opposite in sign.  Thus, for $\nu_\alpha \rightarrow \nu_\alpha$ there is no atmospheric-solar CPC interference, however there is a CPC interference within the atmospheric sector. In eq.~\ref{eq:disapp}, we have given the disappearance probability in such a fashion that this CPC interference  is separated out from the rest of the probability.   For other choices, such as in~\cite{Huber:2019frh}, there is no CPC interference within the atmospheric sector, but there is an atmospheric-solar CPC interference term. \\

The discussions here can easily be carried over to neutrino propagation in uniform matter by using the arguments in the paragraph containing eq.~\ref{eq:CP}. Here the amplitude ${\cal A}(U^*)$ represents the amplitude for anti-neutrino propagation in anti-matter. Thus dividing the probability into terms that is even and odd under the changing the sign of the vacuum Jarlskog invariant. One could call these terms the intrinsic CPC and intrinsic CPV parts. Alternatively one could use T even and odd parts as given in section \ref{sec:TV}. Splitting the intrinsic CPC or T even parts of the amplitude into an atmospheric and solar pieces has similar ambiguities as in vacuum.  \\

In summary, the only uniquely defined interference term in neutrino oscillation is the CPV interference because this term can only arise as the interference between the CP even and CP odd parts of the amplitude, i.e. even and odd under the exchange of $U \leftrightarrow U^*$, eq.~\ref{eq:CP}.  The CPC interference terms are not unique, as parts of the sub-amplitudes can be moved between the atmospheric and solar sub-amplitudes. This affects both the square of the sub-amplitudes and their CPC interference.   
Therefore, the only unique separation of the total oscillation probability is the separation into a CP even and CP odd part with the CP odd part only arising from a CPV interference between the CP even and CP odd parts of the full oscillation amplitude. One of the requirements for this CPV interference to be none zero, is that the amplitude consist of more than one sub-amplitude. e.g. an atmospheric and solar sub-amplitude. However the size of this CPV interference is independent of how one makes the separation between the atmospheric and solar sub-amplitudes and is thus unique.  For the CP even part of the oscillation probability there is no unique way to separate this into squares of sub-amplitudes and CPC interference terms. In conclusion, only the total CPC  and the total CPV parts of the vacuum oscillation probability are unique.\\


\begin{acknowledgments}

GB was supported by the Spanish Grants No.\ PID2020-113334GB-I00/AEI/1013039/501100011033 and No.\ CIPROM/2021/054 (Generalitat Valenciana). SP thanks Ting Cheng and Peter Denton for useful discussions on this manuscript. SP acknowledges support by the United States Department of Energy under Grant Contract No.~DE-AC02-07CH11359. \\

This project has also received support from the European Union's Horizon 2020 research and innovation programme under the Marie  Sklodowska-Curie grant agreement No 860881-HIDDeN as well as under the Marie Skłodowska-Curie Staff Exchange grant agreement No 101086085 - ASYMMETRY.
\end{acknowledgments}

\bibliography{OscAmps}

\appendix

\section{Other ways to write the Appearance Amplitudes}
\label{appx:Amps}

 Without the constraint  $ \xi W_3 \rightarrow 0 \quad \text{ when}  \quad W_3 \rightarrow 0 $ there are other interesting ways to write the neutrino appearance amplitude. If $\xi=(W_2-W_1)/W_3$, we have
\begin{align}
{\cal A}&= 2 W_1 \sin \Delta_{31} e^{-i\Delta_{32}} + 2 W_2\sin \Delta_{32} e^{-i\Delta_{31}}\,, 
\end{align}
which has no 21 or solar sub-amplitude. This is equivalent to removing the $\nu_3$ term in eq.~\ref{eq:amp1}.  All the solar parts come from the interference between 31 and 32 sub-amplitudes which also generates the  CPV interference term. The amplitude, written this way, does not lend itself to being separated into an atmospheric and solar sub-amplitude in any obvious way.\\

Another, is a total symmetric way to write the amplitude, where all sub-amplitudes are treated symmetrically, using $\xi=\frac1{3}(W_2-W_1)/W_3$:
\begin{align}
{\cal A}&=\frac{2}{3} ( W_3-W_1) \sin \Delta_{31} e^{-i\Delta_{32}} \notag \\ &
+ \frac{2}{3} (W_3-W_2)\sin \Delta_{32} e^{-i\Delta_{31}}   \notag \\ & +\frac{2}{3} (W_2- W_1) \sin \Delta_{21} \,.
\end{align}
With this form of the amplitude the interference terms between the sub-amplitudes are particularly significant. \\

We impose the constraint on $\xi$ since we are interested in discussing the interference between the atmospheric and solar amplitudes. \\

\section{Useful IDs}
\label{appx:ids}

The following trigonometric equation is useful to calculate the interference term when squaring eq.~\ref{eq:amp_ap}:
\begin{align}
 \sin ( \Delta_{31} + \Delta_{32} )\sin\Delta_{21} &=  \sin ( \Delta_{31} + \Delta_{32} )\sin ( \Delta_{31} - \Delta_{32} ) \notag \\ 
=& \sin^2 \Delta_{31}  - \sin^2 \Delta_{32}  \,.
\end{align}

The unitarity condition for the $\alpha \beta$ channel $W_1+W_2+W_3=0$, can be used to derive a number of useful identities, for (i,j,k) all different:
\begin{align}
-2R_{ij} &=|W_i|^2 + |W_j|^2 -|W_k|^2  \,, \notag \\
-4 R_{ij} &=  |W_j-W_i|^2 - |W_k|^2 \,, \notag \\
|W_k|^2 &=-R_{ik}-R_{jk} >0 \,.
\end{align}
The last equations implies that for any channel at most only one of $R_{ij}$, $R_{ik}$ and $R_{jk}$ can be positive.

Adding the $\alpha \beta$ and $\alpha \gamma$ channels, $(\alpha, \beta, \gamma)$ all different:
\begin{align}
|U_{\alpha i}|^2 |U_{\alpha j}|^2
&=-R^{\alpha \beta}_{ij} - R^{\alpha \gamma}_{ij} > 0
\end{align}
The implication of this last equation is that, for fixed (i,j), at most only one of $R^{\alpha \beta}_{ij}$, $ R^{\alpha \gamma}_{ij}$ and $ R^{\beta \gamma}_{ij}$ can be positive. See Fig.~\ref{fig:Rplot}.

\end{document}